\documentclass[preprint,aps,amssymb,showpacs]{revtex4}

\usepackage{bm}
\usepackage{graphics,hyperref,textcomp}

\begin{document}

\title{Multiparticle States and the Hadron Spectrum on the Lattice}

\author{David C. Moore}
\author{George T. Fleming}

\affiliation{Sloane Physics Laboratory, Yale University, New Haven,
CT 06520, USA}

\date{\today}

\begin{abstract}
The Clebsch-Gordan decomposition is calculated for direct products
of the irreducible representations of the cubic space group.  These
results are used to identify multiparticle states which appear in
the hadron spectrum on the lattice.  Consideration of the cubic
space group indicates how combinations of both zero momentum and
non-zero momentum multiparticle states contribute to the spectrum.

\end{abstract}


\maketitle

\section{Introduction}

In Lattice QCD with sufficiently light quarks, the only stable
particles are flavored pseudoscalar $(J^P = 0^-)$ mesons and
$J^P=\frac{1}{2}^+$ baryons.  At unphysically heavy quark masses,
the lowest single particle resonance states in other quantum number
channels become stable once they fall below decay thresholds,
\textit{e.g.}\ $\rho \to \pi \pi$ or $\Delta \to p \pi$.
Historically, Lattice QCD has been used to determine the masses of
these stable ground state resonances which are then extrapolated to
physical light quark regime to estimate the physical masses of the
unstable resonances \cite{Butler:1994em,Aoki:1999ff,Aoki:1999yr,
Bowler:1999ae,AliKhan:2001tx,Aoki:2002uc,Namekawa:2004bi}.  To
extract the masses of excited state resonances, or ground state
resonances above threshold, in Lattice QCD is a more involved
process that essentially incorporates the study of multiparticle
scattering states in a finite box
\cite{DeWitt:1956,Huang:1957,Hamber:1983vu,Luscher:1985dn,Luscher:1986pf,
Luscher:1991cf,Rummukainen:1995vs,Kim:2005gf,Christ:2005gi}. See
\cite{Michael:2005kw} for a recent review.

There have been many previous studies in the still nascent field of
multiparticle states in Lattice QCD.  These include $\pi\pi$
scattering
\cite{Sharpe:1992pp,Gupta:1993rn,Kuramashi:1993ka,Fukugita:1994na,
Fukugita:1994ve,Fiebig:1999hs,Liu:2001ss,Aoki:2002in,Aoki:2002ny,
Yamazaki:2004qb,Aoki:2005uf,Beane:2005rj}, $N N$ scattering
\cite{Fukugita:1994na,Fukugita:1994ve,Beane:2006mx}, heavy-light
meson scattering \cite{Cook:2002am}, searching for pentaquark
resonances in $K N$ scattering
\cite{Csikor:2003ng,Sasaki:2003gi,Mathur:2004jr,Chiu:2004gg,Ishii:2004qe,Lasscock:2005tt,
Csikor:2005xb,Alexandrou:2005gc,Takahashi:2005uk,Lasscock:2005kx},
and hadronic decays
\cite{McNeile:2000xx,McNeile:2002fh,McNeile:2002az,Cook:2005az}.

There have also been several studies of excited state resonance
masses in Lattice QCD: Roper resonance and other excited baryons
\cite{Sasaki:2001nf,Edwards:2003cd,Dong:2003br,Mathur:2003zf,
Brommel:2003jm,Chen:2004gp,Burch:2004he,Guadagnoli:2004wm,Sasaki:2005ap,
Burch:2006cc}, charmonium and bottomonium
\cite{Chen:2000ej,Okamoto:2001jb,Liao:2002rj,
Liao:2002,Bernard:2003jd,Gray:2005ur,Luo:2005zg}, heavy-light mesons
\cite{Wingate:2002fh}.  As mentioned above, the masses of all these
resonances lie somewhere within the discrete energy levels of the
multiparticle scattering states and must be disentangled. It is
usually not sufficient to just compute two-point correlation
functions between single particle operators unless it can be clearly
demonstrated that the overlaps of the operators with the
multiparticle states are small enough that only resonances
contribute to the correlator.

In essentially all of the calculations referenced above, the
operators used to compute correlation functions were constructed to
transform irreducibly under the symmetry group of continuum QCD
Hamiltonian.  It is well known \cite{Johnson:1982yq} that these
operators need not transform irreducibly under the symmetry group of
the lattice QCD Hamiltonian.  When calculating ground state masses,
ignoring this fact usually does not lead to confusion.  One possible
exception is the $I(J^P)$ = $\frac{1}{2}(\frac{3}{2}^+)$ and
$\frac{1}{2}(\frac{5}{2}^+)$ baryons, whose lowest-lying resonances
should correspond to the experimentally observed $N(1720)\ P_{13}$
and $N(1680)\ F_{15}$, respectively \cite{Basak:2004hr}.

When calculating properties of single particle resonances at
non-zero momentum, establishing continuum quantum number assignments
will be more difficult than for resonances in the rest frame
\cite{Moore:2005dw,Moore:2006inpub}.  In this work, we demonstrate
for each set of quantum numbers in the center-of-mass frame what
two-particle decompositions are possible, including states with
non-zero relative momentum.

\section{Clebsch-Gordan Decomposition}

We wish to calculate the decomposition of the direct product of
irreducible representations of $\mathcal{T}^3_{lat}\rtimes
\mathrm{O}^\mathrm{D}_h$ into a direct sum of irreducible
representations.  Using the character table for
$\mathcal{T}^3_{lat}\rtimes \mathrm{O}^\mathrm{D}_h$, the character
of a group element $g \in \mathcal{T}^3_{lat}\rtimes
\mathrm{O}^\mathrm{D}_h$ in the direct product representation
$\Gamma_i \otimes \Gamma_j$ is given as $\chi^{\Gamma_i,\Gamma_j}(g)
= \chi^{\Gamma_i}(g) \chi^{\Gamma_j}(g)$. The irreducible
representations $\Gamma_k$ and character table for
$\mathcal{T}^3_{lat}\rtimes \mathrm{O}^\mathrm{D}_h$ are described
in \cite{Moore:2005dw}.  Then, the multiplicity $m$ that an irreducible
representation $\Gamma_k$ is contained in the direct product
representation $\Gamma_i \otimes \Gamma_j$ is given by:
\begin{equation}
\label{eq:proj} m = \frac{1}{|G|} \sum_g \chi^{\Gamma_k}(g)^\ast
\chi^{\Gamma_i,\Gamma_j}(g)
\end{equation}
where the sum is taken over all group elements $g$, and $|G|$ is the
order of the group.  This formula applies to finite lattices, and we
take the limit of $m$ as the lattice size becomes arbitrarily large

The irreducible representations of $\mathcal{T}^3_{lat}\rtimes
\mathrm{O}^\mathrm{D}_h$ are labeled by the magnitude of a lattice
momentum $\mathbf{k}$ and by $\alpha$, which labels an irreducible
representation of the little group of $\mathbf{k}$. The irreducible
representations labeled by lattice vectors $\mathbf{k}$ and
$\mathbf{k'}$ are equivalent if there is a group element $g \in
\mathrm{O}^\mathrm{D}_h$ such that $\mathbf{k'} = g \mathbf{k}$, and
the set of such $\mathbf{k'}$ is called the star of $\mathbf{k}$.
Thus, the inequivalent irreducible representations of
$\mathcal{T}^3_{lat}\rtimes \mathrm{O}^\mathrm{D}_h$ are labeled by
a star, denoted $|\mathbf{k}|$ in analogy with the continuum
notation (we must be careful with vectors such as $\mathbf{k} =
(3,0,0)$ and $\mathbf{k'} = (2,2,1)$ since even though $|\mathbf{k}|
= |\mathbf{k'}|$, they are not in the same star in the discrete
group). As we expect, we see that linear momentum is conserved, i.e.
the product of two representations with momenta $|\mathbf{k_1}|$ and
$|\mathbf{k_2}|$ gives only representations labeled by
$|\mathbf{k}|$ which are the sum of some vector in the star of
$\mathbf{k_1}$ and some vector in the star of $\mathbf{k_2}$. Thus,
the direct product of two irreducible representations of
$\mathcal{T}^3_{lat} \rtimes \mathrm{O}^\mathrm{D}_h$ contain
irreducible representations labeled by $|\mathbf{k}| = 0$ (the
irreducible representations of $\mathrm{O}^\mathrm{D}_h$) if and
only if $|\mathbf{k_1}| = |\mathbf{k_2}|$.  In this paper we are
mainly concerned with such direct products of irreducible
representations which have overlap with zero momentum states.

For the case where $|\mathbf{k_1}| = |\mathbf{k_2}| = 0$, we have
the Clebsch-Gordan decomposition for products of the representations
of $\mathrm{O}^\mathrm{D}_h$ which is given in Tab.~\ref{tab:cgOhd}.
The single valued irreducible representations of
$\mathrm{O}^\mathrm{D}$ are labeled $A_1,A_2,E,T_1,T_2$ and the
double valued representations are labeled $G_1,G_2,H$ using the
Mulliken convention.  The correspondence of these lattice states to
continuum spin states is well-known and given in
Tab.~\ref{tab:subduction} \cite{Johnson:1982yq}. If we include parity to
form the group $\mathrm{O}^\mathrm{D}_h$, then each representation
carries either the label $g$ or $u$ which correspond to positive and
negative parity respectively.  These labels are omitted in the table
since they follow the same multiplication rules that hold in the
continuum: $g \cdot g = u \cdot u = g$ and $g \cdot u = u \cdot g =
u$.

\begin{table*}
\caption{\label{tab:cgOhd}Clebsch-Gordan decomposition for
$\mathrm{O}^\mathrm{D}$.  The table for $\mathrm{O}^\mathrm{D}_h$
adds a parity $g$ or $u$ to each irreducible representation}
\begin{ruledtabular}
\begin{tabular}{lllp{1.8cm}p{1.8cm}p{1.8cm}p{1.8cm}p{1.8cm}p{1.8cm}p{1.8cm}}

$\otimes$ & $A_{1}$ & $A_{2}$ & $E$ & $T_{1}$ & $T_{2}$ & $G_{1}$ &
$G_{2}$
& $H$ \\
\hline $A_{1}$ & $A_{1}$ & $A_{2}$ & $E$ & $T_{1}$ & $T_{2}$ &
$G_{1}$ & $G_{2}$ &
$H$ \\
$A_{2}$ &  & $A_{1}$ & $E$ & $T_{2}$ & $T_{1}$ & $G_{2}$ & $G_{1}$ & $H$ \\
$E$ &   &   & $A_{1} \oplus A_{2} \oplus E$ & $T_{1} \oplus T_{2}$ &
$T_{1} \
\oplus T_{2}$ & $H$ & $H$ & $G_{1} \oplus G_{2} \oplus H$ \\
$T_{1}$ &   &    &   & $A_{1} \oplus E \oplus T_{1} \oplus T_{2}$ &
$A_{2} \oplus E \oplus T_{1} \oplus T_{2}$ & $G_{1} \oplus H$ &
$G_{2} \oplus H$ &
$G_{1} \oplus G_{2} \oplus 2H$ \\
$T_{2}$ &   &    &    &   & $A_{1} \oplus E \oplus T_{1} \oplus
T_{2}$ &
$G_{2} \oplus H$ & $G_{1} \oplus H$ & $G_{1} \oplus G_{2} \oplus 2H$ \\
$G_{1}$ &   &    &    &    &   & $A_{1} \oplus T_{1}$ & $A_{2}
\oplus T_{2}$
& $E \oplus T_{1} \oplus T_{2}$ \\
$G_{2}$ &   &    &    &    &    &   & $A_{1} \oplus T_{1}$ & $E
\oplus T_{1}
\oplus T_{2}$ \\
$H$ &   &    &    &    &    &    &   & $A_{1} \oplus A_{2} \oplus E
\oplus
2T_{1} \oplus 2T_{2}$ \\
\end{tabular}
\end{ruledtabular}
\end{table*}

\begin{table*}
\caption{\label{tab:subduction}The decomposition of continuum SU(2)
spin states $J$ to the representations of $\mathrm{O^D}$ and the
continuum $\mathrm{O}(2)^\mathrm{D}$ states $m_j$ to the various
little groups.}
\begin{ruledtabular}
\begin{tabular}{lp{1.7cm}lllll}

$J/m_j$ & $\mathrm{O^D}$ & $\mathrm{Dic}_4$ & $\mathrm{Dic}_3$ & $\mathrm{Dic}_2$ & $\mathrm{C_{4}}$ & $\mathrm{C_{2}}$ \\
\hline

  $0^+$         & $A_{1g}$& $A_1$            & $A_1$            & $A_1$            & $A_1$ & $A$ \\
  $0^-$         & $A_{1u}$& $A_2$            & $A_2$            & $A_2$            & $A_2$ & $A$ \\
  $\frac{1}{2}$ &$G_1$& $E_1$            & $E_1$            & $E$              & $E$ & $2B$ \\
  $1$           & $T_1$& $E_2$            & $E_2$            & $B_1 \oplus B_2$ & $A_1 \oplus A_2$ & $2A$ \\
  $\frac{3}{2}$ &  $H$& $E_3$            & $B_1 \oplus B_2$ & $E$              & $E$ & $2B$ \\
  $2$           & $E \oplus T_2$& $B_1 \oplus B_2$ & $E_2$            & $A_1 \oplus A_2$ & $A_1 \oplus A_2$ & $2A$ \\
  $\frac{5}{2}$ & $G_2 \oplus H$& $E_3$            & $E_1$            & $E$              & $E$ & $2B$ \\
  $3$           & $A_2 \oplus T_1 \oplus T_2$& $E_2$            & $A_1 \oplus A_2$ & $B_1 \oplus B_2$ & $A_1 \oplus A_2$ & $2A$ \\
  $\frac{7}{2}$ & $G_1 \oplus G_2 \oplus H$& $E_1$            & $E_1$            & $E$              & $E$ & $2B$ \\
  $4$           & $A_1 \oplus E \oplus T_1 \oplus T_2$ & $A_1 \oplus A_2$ & $E_2$            & $A_1 \oplus A_2$ & $A_1 \oplus A_2$ & $2A$ \\

\end{tabular}
\end{ruledtabular}
\end{table*}

\section{Multiparticle States}
From Tabs.~\ref{tab:cgOhd} and \ref{tab:subduction}, we see that
multiparticle states for the lowest energy states for each spin
behave as in the continuum.  In fact, we can generate much of
Tab.~\ref{tab:cgOhd} using Tab.~\ref{tab:subduction} and the
continuum rules for addition of angular momentum.  For example,
continuum spins $1 \otimes 2 = 1 \oplus 2 \oplus 3$ and the
corresponding lattice representations $T_1 \otimes (E \oplus T_2) =
(T_1 \oplus T_2) \oplus (A_2 \oplus E \oplus T_1 \oplus T_2) = T_1
\oplus (E \oplus T_2) \oplus (A_2 \oplus T_1 \oplus T_2)$.  The
continuum relations hold, of course, because taking the direct sum
of the lattice irreducible representations in
Tab.~\ref{tab:subduction} gives equivalent representations to the
SU(2) irreducible representations, so they must follow  the same
multiplication rules. For spins which lie in a single lattice
irreducible representation, there is no ambiguity in the
combinations.  Thus, on the lattice, the combination of low spin
single particle states is identical to the continuum. The $\pi$ with
$J^P = 0^-$ lies in the irreducible representation $A_{1u}$, and the
$\pi\pi$ multiparticle state lies in $A_{1u} \otimes A_{1u} =
A_{1g}$, which as expected corresponds to $J^P = 0^+$.  Similarly,
for the vector meson $\rho(770)$ which lies in $T_{1u}$, then the
state $\rho\pi$ lies in $T_{1g}$ which corresponds to $J^P = 1^+$.

Higher spin states are less straightforward since multiple lattice
irreducible representations appear for each spin. A spin 2 continuum
state could lie in either the $E$ or $T_2$ representations on the
lattice, which leads to different possibilities for multiparticle
states.  For the continuum example $1 \otimes 2$, then if the spin 2
state lies in $E$, the combined state is $T_1 \oplus T_2$ whereas if
it were in $T_2$, the combined state would be $A_2 \oplus E \oplus
T_1 \oplus T_2$. In both cases, we get combinations of the
representations $A_2, E, T_1$, and $T_2$, all of which have their
lowest spins corresponding to 1,2, or 3, but the identification of a
particular spin state to each is difficult without further
information.

When combining states with non-zero momentum, the continuum
relations are not as easily recovered.
Tabs.~\ref{tab:cgn00}~-~\ref{tab:cgnnm} show the zero-momentum
representations in the decomposition of products of irreducible
representations labeled by the possible lattice momenta.  Here, the
continuum representations are no longer labeled by $J$, but by the
projection of $J$ along the momentum vector, $m_j$.  However, we
know that a particle with a given $m_j$ has $J \geq m_j$, so the
lowest spin state for a given irreducible representation of
$\mathcal{T}^3_{lat}\rtimes \mathrm{O}^\mathrm{D}_h$ will be $J =
m_j$. In addition, the reduced symmetry of the little groups leads
to fewer distinct lattice irreducible representations than at zero
momentum.  Since the continuum irreducible representations are
mapped to fewer lattice representations, it is more difficult to
assign a particular spin to a given lattice irreducible
representation.

Thus, for low spins on the lattice we expect to see the same
multiparticle states as we would in the continuum, but as we go to
higher spins or non-zero momentum deviations will occur from the
continuum behavior.  For example, from Tab.~\ref{tab:subduction} a
$J^P = 2^+$ state in the continuum can lie in either $E_g$,
$T_{2g}$, or some combination of both on the lattice. In the
continuum, an $f_2$(1270) meson with $J^P = 2^+$ has the decay modes
$\pi\pi$, $4\pi$, and $K\bar{K}$ \cite{PDBook}. Thus we expect to
see multiparticle states in the lattice spectrum corresponding to
these decay modes. From Tab.~\ref{tab:cgn00}, we see that the
multiparticle state $\{n,0,0\}; A_2 \otimes \{n,0,0\}; A_2$
corresponding to these decay modes occurs in the $E_g$ channel, but
not in the $T_{2g}$ channel.  We must calculate exactly how the
particular spin 2 continuum state we are interested in subduces to
the lattice to determine whether these multiparticle states will
appear. In this case, it is possible that states we expect from the
continuum rules for addition of angular momentum would be absent
from the lattice spectrum.

Another example where the lattice multiparticle states cannot be
predicted from the continuum behavior occurs for $J^P =
\frac{5}{2}^-$.  As for spin 2, a continuum spin $\frac{5}{2}$ state
can lie in more than one lattice representation, either $G_{2u}$ or
$H_u$.  However, if we consider multiparticle states in the $H_u$
channel, then both states which go to spin $\frac{3}{2}$ in the
continuum limit and states which go to spin $\frac{5}{2}$ in the
continuum limit will appear, since at any finite lattice spacing
these states may have the same lattice quantum numbers (i.e.
correspond to the $H_u$ irreducible representation).  Again, we must
know how our $J^P = \frac{5}{2}$ state subduces to the lattice in
order to determine exactly which multiparticle states will occur in
the lattice spectrum.  Here, multiparticle states are present in the
lattice spectrum which we would not predict from the continuum
states.  As we consider higher spins, these types of ambiguities
become common.

\begin{table*}
\begin{ruledtabular}
\caption{\label{tab:cgn00}Clebsch-Gordan decomposition for products
of the irreducible representations of $\mathcal{T}^3_{lat}\rtimes
\mathrm{O}^\mathrm{D}_h$ labeled by $|(n,0,0)|$. Only the zero
momentum representations are given.}

\begin{tabular}{lp{2.0cm}p{2.0cm}p{2.0cm}p{2.0cm}p{2.0cm}p{2.0cm}p{2.0cm}}
$\otimes$ & $A_1$ & $A_2$ & $B_1$ & \
$B_2$ & $E_1$ & $E_2$ & $E_3$ \\
\hline

$A_1$ & $A_{1g} \oplus E_{g} \oplus T_{1u}$ & $T_{1g} \oplus A_{1u}
\oplus E_{u}$ & $A_{2g} \oplus E_{g} \oplus T_{2u}$ & $T_{2g} \oplus
 A_{2u} \oplus E_{u}$ & $G_{1g} \oplus H_{g} \oplus G_{1u} \oplus
H_{u}$ & $T_{1g} \oplus T_{2g} \oplus T_{1u} \oplus T_{2u}$ &
$G_{2g} \oplus H_{g}
\oplus G_{2u} \oplus H_{u}$ \\
$A_2$ &  & $A_{1g} \oplus E_{g} \oplus T_{1u}$ & $T_{2g} \oplus
A_{2u} \oplus E_{u}$ & $A_{2g} \oplus E_{g} \oplus T_{2u}$ & $G_{1g}
\oplus  H_{g} \oplus G_{1u} \oplus H_{u}$ & $T_{1g} \oplus T_{2g}
\oplus T_{1u}
\oplus T_{2u}$ & $G_{2g} \oplus H_{g} \oplus G_{2u} \oplus H_{u}$ \\
$B_1$ &   &   & $A_{1g} \oplus E_{g} \oplus T_{1u}$ & $T_{1g} \oplus
A_{1u} \oplus E_{u}$ & $G_{2g} \oplus H_{g} \oplus G_{2u} \oplus
H_{u}$ & $T_{1g} \oplus T_{2g} \oplus T_{1u} \oplus T_{2u}$ &
$G_{1g} \oplus
H_{g} \oplus G_{1u} \oplus H_{u}$ \\
$B_2$ &   &    &   & $A_{1g} \oplus E_{g} \oplus T_{1u}$ & $G_{2g}
\oplus H_{g} \oplus G_{2u} \oplus H_{u}$ & $T_{1g} \oplus T_{2g}
\oplus T_{1u} \oplus T_{2u}$ & $G_{1g} \oplus H_{g} \oplus G_{1u}
\oplus
H_{u}$ \\
$E_1$ &   &    &    &   & $A_{1g} \oplus E_{g} \oplus 2T_{1g} \oplus
T_{2g} \oplus A_{1u} \oplus E_{u} \oplus 2T_{1u} \oplus T_{2u}$ &
$G_{1g} \oplus G_{2g} \oplus 2H_{g} \oplus G_{1u} \oplus G_{2u}
\oplus 2H_{u}$ & $A_{2g} \oplus E_{g} \oplus T_{1g} \oplus 2T_{2g}
\oplus A_{2u}
\oplus E_{u} \oplus T_{1u} \oplus 2T_{2u}$ \\
$E_2$ &   &    &    &    &   & $A_{1g} \oplus A_{2g} \oplus 2E_{g}
\oplus T_{1g} \oplus T_{2g} \oplus A_{1u} \oplus A_{2u} \oplus
2E_{u} \oplus T_{1u} \oplus T_{2u}$ & $G_{1g} \oplus G_{2g} \oplus
2H_{g} \oplus
G_{1u} \oplus G_{2u} \oplus 2H_{u}$ \\
$E_3$ &   &    &    &    &    &   & $A_{1g} \oplus E_{g} \oplus
2T_{1g} \oplus T_{2g} \oplus A_{1u} \oplus E_{u} \oplus 2T_{1u}
\oplus
T_{2u}$ \\

\end{tabular}
\end{ruledtabular}
\end{table*}

\begin{table*}
\caption{\label{tab:cgnn0}Zero-momentum irreducible representations
for $|(n,n,0)|$.}
\begin{ruledtabular}
\begin{tabular}{lp{2.6cm}p{2.6cm}p{2.6cm}p{2.6cm}p{2.6cm}}
$\otimes$ & $A_1$ & $A_2$ & $B_1$ & \
$B_2$ & $E$ \\
\hline

$A_1$ & $A_{1g} \oplus E_{g} \oplus T_{2g} \oplus T_{1u} \oplus \
T_{2u}$ & $T_{1g} \oplus T_{2g} \oplus A_{1u} \oplus E_{u} \oplus
T_{2u}$ & $A_{2g} \oplus E_{g} \oplus T_{1g} \oplus T_{1u} \oplus
T_{2u}$ & $T_{1g} \oplus T_{2g} \oplus A_{2u} \oplus E_{u} \oplus
T_{1u}$ & $G_{1g} \oplus \
G_{2g} \oplus 2H_{g} \oplus G_{1u} \oplus G_{2u} \oplus 2H_{u}$ \\
$A_2$ &  & $A_{1g} \oplus E_{g} \oplus T_{2g} \oplus T_{1u} \oplus
T_{2u}$ & $T_{1g} \oplus T_{2g} \oplus A_{2u} \oplus E_{u} \oplus \
T_{1u}$ & $A_{2g} \oplus E_{g} \oplus T_{1g} \oplus T_{1u} \oplus
T_{2u}$ & $G_{1g} \oplus G_{2g} \oplus 2H_{g} \oplus G_{1u} \oplus
G_{2u} \oplus \
2H_{u}$ \\
$B_1$ &   &   & $A_{1g} \oplus E_{g} \oplus T_{2g} \oplus T_{1u} \
\oplus T_{2u}$ & $T_{1g} \oplus T_{2g} \oplus A_{1u} \oplus E_{u}
\oplus T_{2u}$ & $G_{1g} \oplus G_{2g} \oplus 2H_{g} \oplus G_{1u}
\oplus G_{2u} \
\oplus 2H_{u}$ \\
$B_2$ &   &    &   & $A_{1g} \oplus E_{g} \oplus T_{2g} \oplus \
T_{1u} \oplus T_{2u}$ & $G_{1g} \oplus G_{2g} \oplus 2H_{g} \oplus
G_{1u} \
\oplus G_{2u} \oplus 2H_{u}$ \\
$E$ &   &    &    &   & $A_{1g} \oplus A_{2g} \oplus 2E_{g} \oplus
3T_{1g} \oplus 3T_{2g} \oplus A_{1u} \oplus A_{2u} \oplus 2E_{u} \
\oplus 3T_{1u} \oplus 3T_{2u}$ \\

\end{tabular}
\end{ruledtabular}
\end{table*}

\begin{table*}
\caption{\label{tab:cgnnn}Zero-momentum irreducible representations
for $|(n,n,n)|$.}
\begin{ruledtabular}
\begin{tabular}{lp{2.2cm}p{2.2cm}p{2.2cm}p{2.2cm}p{2.2cm}p{2.2cm}}
$\otimes$ & $A_1$ & $A_2$ & $E_2$ & \
$B_1$ & $B_2$ & $E_1$ \\
\hline

$A_1$ & $A_{1g} \oplus T_{2g} \oplus A_{2u} \oplus T_{1u}$ & \
$A_{2g} \oplus T_{1g} \oplus A_{1u} \oplus T_{2u}$ & $E_{g} \oplus
T_{1g} \oplus T_{2g} \oplus E_{u} \oplus T_{1u} \oplus T_{2u}$ &
$H_{g} \oplus  H_{u}$ & $H_{g} \oplus H_{u}$ & $G_{1g} \oplus G_{2g}
\oplus H_{g} \oplus
G_{1u} \oplus G_{2u} \oplus H_{u}$ \\
$A_2$ &  & $A_{1g} \oplus T_{2g} \oplus A_{2u} \oplus T_{1u}$ &
$E_{g} \oplus T_{1g} \oplus T_{2g} \oplus E_{u} \oplus T_{1u} \oplus
T_{2u}$  & $H_{g} \oplus H_{u}$ & $H_{g} \oplus H_{u}$ & $G_{1g}
\oplus G_{2g} \oplus
H_{g} \oplus G_{1u} \oplus G_{2u} \oplus H_{u}$ \\
$E_2$ &   &   & $A_{1g} \oplus A_{2g} \oplus E_{g} \oplus 2T_{1g}
\oplus 2T_{2g} \oplus A_{1u} \oplus A_{2u} \oplus E_{u} \oplus
2T_{1u}  \oplus 2T_{2u}$ & $G_{1g} \oplus G_{2g} \oplus H_{g} \oplus
G_{1u} \oplus \ G_{2u} \oplus H_{u}$ & $G_{1g} \oplus G_{2g} \oplus
H_{g} \oplus G_{1u} \ \oplus G_{2u} \oplus H_{u}$ & $G_{1g} \oplus
G_{2g} \oplus 3H_{g} \oplus
G_{1u} \oplus G_{2u} \oplus 3H_{u}$ \\
$B_1$ &   &    &   & $A_{2g} \oplus T_{1g} \oplus A_{1u} \oplus
T_{2u}$ & $A_{1g} \oplus T_{2g} \oplus A_{2u} \oplus T_{1u}$ &
$E_{g} \oplus
T_{1g} \oplus T_{2g} \oplus E_{u} \oplus T_{1u} \oplus T_{2u}$ \\
$B_2$ &   &    &    &   & $A_{2g} \oplus T_{1g} \oplus A_{1u} \oplus
T_{2u}$ & $E_{g} \oplus T_{1g} \oplus T_{2g} \oplus E_{u} \oplus
T_{1u} \oplus T_{2u}$ \\
$E_1$ &   &    &    &    &   & $A_{1g} \oplus A_{2g} \oplus  E_{g}
\oplus 2T_{1g} \oplus 2T_{2g} \oplus A_{1u} \oplus A_{2u} \oplus
E_{u}
\oplus 2T_{1u} \oplus 2T_{2u}$ \\

\end{tabular}
\end{ruledtabular}
\end{table*}

\begin{table*}
\caption{\label{tab:cgnm0}Zero-momentum irreducible representations
for $|(n,m,0)|$.}
\begin{ruledtabular}
\begin{tabular}{lp{4.2cm}p{4.2cm}p{4.2cm}}
$\otimes$ & $A_1$ & $A_2$ & $E$ \\
\hline

$A_1$ & $A_{1g} \oplus A_{2g} \oplus 2E_{g} \oplus T_{1g} \oplus
T_{2g} \oplus 2T_{1u} \oplus 2T_{2u}$ & $2T_{1g} \oplus 2T_{2g}
\oplus A_{1u} \oplus A_{2u} \oplus 2E_{u} \oplus T_{1u} \oplus
T_{2u}$ & $2G_{1g} \oplus 2G_{2g} \oplus 4H_{g} \oplus
2G_{1u} \oplus 2G_{2u} \oplus 4H_{u}$\\
$A_2$ &  & $A_{1g} \oplus A_{2g} \oplus 2E_{g} \oplus T_{1g} \oplus
T_{2g} \oplus 2T_{1u} \oplus 2T_{2u}$ & $2G_{1g} \oplus
2G_{2g} \oplus 4H_{g} \oplus 2G_{1u} \oplus 2G_{2u} \oplus 4H_{u}$ \\
$E$ &   &   & $2A_{1g} \oplus 2A_{2g} \oplus 4E_{g} \oplus 6T_{1g}
\oplus 6T_{2g} \oplus 2A_{1u} \oplus 2A_{2u} \oplus 4E_{u} \oplus
6T_{1u} \oplus 6T_{2u}$\\
\end{tabular}
\end{ruledtabular}
\end{table*}

\begin{table*}
\caption{\label{tab:cgnnm}Zero-momentum irreducible representations
for $|(n,n,m)|$.}
\begin{ruledtabular}
\begin{tabular}{lp{4.2cm}p{4.2cm}p{4.2cm}}
$\otimes$ & $A_1$ & $A_2$ & $E$ \\
\hline

$A_1$ & $A_{1g} \oplus E_{g} \oplus T_{1g} \oplus 2T_{2g} \oplus
A_{2u} \oplus E_{u} \oplus 2T_{1u} \oplus T_{2u}$ & $A_{2g} \oplus
E_{g} \oplus 2T_{1g} \oplus T_{2g} \oplus A_{1u} \oplus E_{u} \oplus
T_{1u} \oplus 2T_{2u}$ & $2G_{1g} \oplus 2G_{2g} \oplus 4H_{g}
\oplus
2G_{1u} \oplus 2G_{2u} \oplus 4H_{u}$\\
$A_2$ &  & $A_{1g} \oplus E_{g} \oplus T_{1g} \oplus 2T_{2g} \oplus
A_{2u} \oplus E_{u} \oplus 2T_{1u} \oplus T_{2u}$ & $2G_{1g} \oplus
2G_{2g} \oplus 4H_{g} \oplus 2G_{1u} \oplus 2G_{2u} \oplus 4H_{u}$ \\
$E$ &   &   & $2A_{1g} \oplus 2A_{2g} \oplus 4E_{g} \oplus 6T_{1g}
\oplus 6T_{2g} \oplus 2A_{1u} \oplus 2A_{2u} \oplus 4E_{u} \oplus
6T_{1u} \oplus 6T_{2u}$\\
\end{tabular}
\end{ruledtabular}
\end{table*}

\section{Conclusion}

We have calculated the Clebsch-Gordan decomposition for the cubic
space group, $\mathcal{T}^3_{lat}\rtimes \mathrm{O}^\mathrm{D}_h$,
which determines the allowed multiparticle states for each lattice
irreducible representation, including states with non-zero momentum.
For states with low spin, we find that the lattice states mirror the
continuum behavior since these continuum irreducible representations
remain irreducible on the lattice.

For higher spins or non-zero momentum, the continuum relations are
not as easily recovered.  Since multiple continuum spins lie in each
lattice irreducible representation, multiparticle states appear
which we would not expect from the continuum behavior.  In the
continuum limit, we should recover the correct multiparticle states,
but at any finite lattice spacing these effects will play a role. In
general, continuum states must be subduced to the lattice
irreducible representations in order to correctly predict which
multiparticle states will appear on the lattice.

\section{Acknowledgments}

\bibliography{cgpaper}
\bibliographystyle{apsrev}

\end{document}